\documentclass[a4paper,12pt]{article}

\usepackage[usenames]{color}
\usepackage{cite}

\usepackage{graphicx}
\usepackage{amsmath}
\usepackage{amssymb}

\begin{document}
\begin{titlepage}
\null
\begin{flushright}
TIT/HEP-590\\
HIP-2008-32/TH\\
September, 2008
\end{flushright}

\vskip 1.8cm
\begin{center}
 
  {\Large \bf Time-dependent and Non-BPS Solutions in $\mathcal{N} = 6$ Superconformal Chern-Simons Theory 
  }

\vskip 2.3cm
\normalsize

  {\bf Toshiaki Fujimori$^{\dagger}$\footnote{fujimori(at)th.phys.titech.ac.jp}, 
Koh Iwasaki$^\dagger$\footnote{iwasaki(at)th.phys.titech.ac.jp}, 
Yoshishige Kobayashi$^\dagger$\footnote{yosh(at)th.phys.titech.ac.jp } 
and Shin Sasaki$^\ddagger$\footnote{shin.sasaki(at)helsinki.fi}}

\vskip 0.5cm

  { \it 
  $^\dagger$Department of Physics, Tokyo Institute of Technology \\
  Tokyo 152-8551, JAPAN \\
  \vskip 0.5cm
  $^\ddagger$Department of Physics,
  University of Helsinki \\
  Helsinki Institute of Physics,
  P.O.Box 64, FIN-00014, Finland \\
  }

\vskip 3cm

\begin{abstract}
We study a class of classical solutions of three-dimensional $\mathcal{N} = 6$ 
superconformal Chern-Simons theory coupled with $U(N) \times U(N)$ 
bi-fundamental matter fields. Especially, time evolutions of fuzzy spheres 
are discussed for both massless and massive cases.
For the massive case, there are a variety of solutions having different 
behaviors according to the value of the mass.
In addition to the time-dependent solutions, we analyze non-BPS static solutions which represent parallel M5/M5 or M5/anti-M5-branes suspended by multiple M2-branes.
These solutions are similar to the fundamental strings connecting two parallel 
(anti) D$p$-branes in perturbative string theory.
A moving M5-brane and domain wall solutions with constant velocity 
that are obtained by the Lorentz boost of the known BPS solutions are briefly addressed.
\end{abstract}

\end{center}
\end{titlepage}

\newpage

\section{Introduction}
Low energy world-volume theory of multiple M2-branes recently attracted much attention
after the proposal of the Bagger-Lambert-Gustavsson (BLG) model \cite{BaLa, Gu}.
In the BLG model, the theory consists of matter fields with 
Chern-Simons term of a novel 3-algebra gauge field. 
This model exhibits an explicit $\mathcal{N} = 8$ supersymmetry with 
manifest $SO(8)$ R-symmetry.
Quantization condition of the 3-algebra structure constant was pointed out in 
\cite{BaLa} which defines the level of the Chern-Simons term.
The explicit Lagrangian was constructed by focusing on 
a 3-algebra with structure constant $f^{abcd} = \epsilon^{abcd}$ known 
as $\mathcal{A}_4$ algebra.
This theory was expected to describe two coincident M2-branes 
located on the ``M-fold'' \cite{DiMuPaRa} and 
various properties of this model have been explored 
\cite{BLG}.
One would expect that higher rank 3-algebras describe more number 
of M2-branes.
However, under the assumption of positive definiteness of the metric, 
there is only one non-trivial example of the 3-algebra, just only
$\mathcal{A}_4$ and its direct product are allowed \cite{Pa, GaGu}.
In order to circumvent this no-go theorem, 3-algebras including Lorentzian 
metric \cite{HoImMa, GoMiRu, BeRoToVe}, non-antisymmetric structure 
constant \cite{BaLa2, ChSa} 
were discussed.
It was addressed that the $\mathcal{A}_4$ algebra can be interpreted as $SO(4)$ gauge symmetry.
This fact provides a valuable intuition for the ordinary gauge group 
description of the model. Indeed, in \cite{Ra}, the $SO(4)$ BLG model were 
re-formulated as an $SU(2) \times SU(2)$ superconformal Chern-Simons-matter theory. 

Meanwhile, Aharony-Bergman-Jafferis-Maldacena recently proposed 
three dimensional 
$\mathcal{N} = 6$ superconformal Chern-Simons-matter theory as an 
alternative model of $N$ coincident 
M2-branes in $\mathbf{C}^4/\mathbf{Z}_k$ orbifold (ABJM model) \cite{AhBeJaMa}. 
The gauge group is $U(N) \times U(N)$ and four complex scalar fields are 
introduced as the (anti) bi-fundamental representation of 
this gauge group. Two gauge fields $A_{\mu}$ and $\hat{A}_{\mu}$ 
corresponding to each $U(N)$s have levels $k$ and $-k$ respectively and 
the Lagrangian of the ABJM model exhibits an $SU(4)$ R-symmetry. 
For $SU(2) \times SU(2)$ gauge group, the $\mathcal{N} = 6$ 
supersymmetry is enhanced to $\mathcal{N} = 8$ and the ABJM model precisely 
recovers the BLG model. It is also argued in \cite{AhBeJaMa} that 
the model is dual to M-theory on $AdS_4 \times S^7/\mathbf{Z}_k$ at large-$N$.
A lot of works on this model have been studied \cite{ABJM}. 

On the other hand, as in the case of the 
world-volume description of D-brane configurations in perturbative 
string theory, 
the classical solutions in the M2-brane world-volume theory 
captures various properties of branes existing in M-theory.
Remarkably, BPS solutions of the ABJM model have been analyzed by several 
authors \cite{Te, HaLi}. In these papers, a fuzzy funnel solution which 
preserves at least $\mathcal{N} = 3$ supersymmetry was 
found. Similar to the Basu-Harvey analysis \cite{BaHa}, the cross-section of this fuzzy 
funnel is fuzzy $S^3$ representing $N$ M2-branes attached to an M5-brane. 
These BPS solutions represent static configurations of the branes and 
are unable to capture the dynamics of brane systems. 
In addition to these static solutions, it is interesting to study time-dependent and non-BPS configurations of branes in M-theory. 

Motivated by these facts, in this paper, we 
demonstrate that there are classical solutions in the ABJM model that represent time evolution of M2-branes or M2/M5-branes combined system. 
In order to study the time dependence of the M2-brane configuration, 
we examine its second differential equation of motion.
In general, it is difficult to solve these non-linear second derivative 
differential equations, however, for some special situation, we can solve 
it analytically.
In addition to these time-dependent solutions, we study static 
solutions that would be interpreted as M2-branes stretched between two 
M5s or M5/anti-M5-branes. 
Similar to the D3$\perp$D1 system \cite{CaMa, CoMyTa, Ha}, 
we find that there are two types of solutions. 
One is a wormhole-like solution, the other is 
a cusp solution. Similar analysis were performed for D5$\perp$D1 
\cite{CoMyTa2} and D7$\perp$D1 \cite{CoKoMu} systems.

The organization of this paper is as follows. 
In section 2, we briefly review the ABJM model and its half-BPS solutions.
The equation of motion is also determined.
In section 3, time-dependent oscillating fuzzy $S^3$ solutions are analyzed 
both for massless and massive cases.
In section 4, non-BPS configurations representing multiple M2-branes 
suspended between parallel M5/anti-M5 and M5/M5 are discussed.
Section 5 is our conclusion and discussion. 
Properties of the BPS matrices can be found in appendix A.
In appendix B, moving fuzzy funnels and domain wall solutions with constant 
velocity are briefly discussed. 
In appendix C, a simple form of the time-dependent solution is presented.

\section{Equations of motion and BPS solutions}
In this section, we briefly review $(2+1)$ dimensional $\mathcal{N} = 6$ 
Chern-Simons theory proposed in \cite{AhBeJaMa}.
This theory consists of $U(N) \times U(N)$ gauge fields with level $k$ 
and $-k$ coupled to (anti) bi-fundamental matter fields.
We basically employ the notation of \cite{BeKlKlSm} but with 
different normalization of $U(N)$ generators $T^a$ such that 
$\mathrm{Tr} (T^a T^b) = \frac{1}{2} \delta^{ab}$. 
The bosonic part of the ABJM action is 
\begin{eqnarray}
S = S_{\mathrm{kin}} + S_{\mathrm{CS}} + S_{\mathrm{pot}}.
\end{eqnarray}
Here each parts are given by
\begin{eqnarray}
S_{\mathrm{kin}} &=& \int \! d^3 x \ 
\mathrm{Tr} 
\left[
- (D_{\mu} Z^A) (D^{\mu} Z^A)^{\dagger} 
- (D_{\mu} W_A) (D^{\mu} W_A)^{\dagger}
\right],
\end{eqnarray}
\begin{eqnarray}
S_{\mathrm{CS}} = 
\frac{k}{4 \pi} \int \! d^3 x \
\mathrm{Tr} \epsilon^{\mu \nu \lambda}
\left[
A_{\mu} \partial_{\nu} A_{\lambda} + \frac{2i}{3} A_{\mu} A_{\nu} A_{\lambda}
- \hat{A}_{\mu} \partial_{\nu} \hat{A}_{\lambda} - \frac{2i}{3} \hat{A}_{\mu} \hat{A}_{\nu} \hat{A}_{\lambda}
\right],
\end{eqnarray}
\begin{eqnarray}
S_{\mathrm{pot}} &=& - \frac{4\pi^2}{k^2} \int \! d^3 x \ 
\mathrm{Tr}
\left[
(Z^A Z^{\dagger}_A + W^{\dagger A} W_A) (Z^B Z^{\dagger}_B - W^{\dagger 
B} W_B) (Z^C Z^{\dagger}_C - W^{\dagger C} W_{C}) \right. \nonumber \\
& & \qquad \qquad \qquad \quad + (Z^{\dagger}_A Z^A + W_A W^{\dagger A}) (Z^{\dagger}_B Z^B - W_B 
W^{\dagger B}) (Z^{\dagger}_C Z^C - W_C W^{\dagger C}) \nonumber \\
& & \qquad \qquad \qquad \quad - 2 Z^{\dagger}_A (Z^B Z^{\dagger}_B - W^{\dagger B} W_B) Z^A 
(Z^{\dagger}_C Z^C - W_C W^{\dagger C}) \nonumber \\
& & \left.
\qquad \qquad \qquad \quad - 2 W^{\dagger A} (Z^{\dagger}_B Z^B - W_B W^{\dagger B}) W_A 
(Z^C Z^{\dagger}_C - W^{\dagger C} W_C)
\right] \nonumber \\
& & + \frac{16\pi^2}{k^2} \int \! d^3 x \ 
\mathrm{Tr}
\left[
W^{\dagger A} Z^{\dagger}_B W^{\dagger C} W_A Z^B W_C - W^{\dagger A} 
Z^{\dagger}_B W^{\dagger C} W_C Z^B W_A \right. \nonumber \\
& & \qquad \qquad \qquad \quad \left. + Z^{\dagger}_A W^{\dagger B} Z^{\dagger}_C Z^A W_B Z^C
- Z^{\dagger}_A W^{\dagger B} Z^{\dagger}_C Z^C W_B Z^A
\right].
\end{eqnarray}
The world-volume metric is chosen such as $\eta_{\mu \nu} = \mathrm{diag} (-1,1,1)$.
Here $A_{\mu}$ and $\hat{A}_{\mu}$ are $U(N) \times U(N)$ gauge fields, 
$Z^A, W^{\dagger A} (A=1,2)$ are bi-fundamental $(\mathbf{N}, 
\bar{\mathbf{N}})$ representation of the $U(N) \times U(N)$ gauge group, 
$k$ is an integer specifying the level of the Chern-Simons theory.
The gauge covariant derivative is 
\begin{eqnarray}
& & D_{\mu} Z^A \equiv \partial_{\mu} Z^A + i A_{\mu} Z^A - i Z^A 
\hat{A}_{\mu}, \nonumber \\
& & D_{\mu} W_A \equiv \partial_{\mu} W_A - i W_A A_{\mu}  + i \hat{A}_{\mu} W_A ,
\end{eqnarray}
and the field strength for $A_{\mu}$ is defined by 
\begin{eqnarray}
F_{\mu \nu} = \partial_{\mu} A_{\nu} - \partial_{\nu} A_{\mu}
+ i [A_{\mu}, A_{\nu}],
\end{eqnarray}
and similarly for $\hat{A}_{\mu}$. 
We normalized overall $U(1)$ charges to be $+1$. 
This model exhibits a manifest $SU(2) \times SU(2) \times U(1)_R$ global symmetry which 
is in fact combined with the $SU(2)_R$ and enhanced to $SU(4)_R$.
For $k>2$, this model has explicit $\mathcal{N} = 6$ supersymmetry. 
The supersymmetry transformation of the component fields can be found in 
\cite{Te, GaGiYi, HoLeLeLePa}. 
This model is expected to describe low energy effective theory of 
$N$ coincident M2-branes in $\mathbf{C}^4/\mathbf{Z}_k$. 
The world-volume coordinates $(t,x^1,x^2)$ are identified with the space-time coordinates $(X^0,X^1,X^2)$.

Once we drop the gauge fields and $W_A$, the BPS equation 
for $Z^A = Z^A (x_2)$, which preserves a half of the original supersymmetry can be derived by performing the Bogomol'nyi completion 
\cite{HaLi}\footnote{In fact, the BPS equation presented here is 
coming from the D-term completion \cite{HaLi}.
The F-term condition in our ansatz provides a trivial solution only.} or 
supersymmetry transformation \cite{Te} as 
\begin{eqnarray}
\partial_2 Z^A = - \frac{2\pi}{k} 
\left(
Z^B Z^{\dagger}_B Z^A - Z^A Z^{\dagger}_B Z^B
\right).
\label{massless_BPS}
\end{eqnarray}
Let us consider an ansatz
\begin{eqnarray}
Z^A = f(x_2) S^A,
\end{eqnarray}
where $S^A$ are constant matrices satisfying the relation 
\begin{eqnarray}
S^A = S^B S^{\dagger}_B S^A - S^A S^{\dagger}_B S^B.
\label{BPS_matrix_condition}
\end{eqnarray}
We call these $S^A$ ``the BPS matrices". 
The matrices were first found in 
\cite{GoRoRaVe} and the explicit form are presented in the appendix A.
The two complex scalar fields $Z^A \ (A = 1,2)$ have 
the physical meaning of the transverse displacement of M2-branes along four-directions, say, $(X^3, X^4, X^7, X^8)$ directions.
If we assume real function $f$ and using these BPS matrices, the BPS 
equation (\ref{massless_BPS}) reduces to 
\begin{eqnarray}
\partial_2 f (x_2) = - \frac{2 \pi}{k} f^3 (x_2).
\label{eom_scalar}
\end{eqnarray}
A solution to this equation is easily found to be 
\begin{eqnarray}
f (x_2) = \sqrt{\frac{k}{4 \pi}} (x_2 - x_0)^{-\frac{1}{2}}, \ x_2 > x_0.
\end{eqnarray}
Because of the noncommutative property of $Z^A$, the solution exhibits 
a fuzzy configuration. Indeed, 
this is just the fuzzy funnel solution with its cross-section fuzzy $S^3$.
This solution represents an M5-brane located at $x_2 = x_0$ extending along $(X^1, X^3, X^4, X^7, X^8)$ and sharing the $X^1$ direction with $N$ M2-branes.

On the other hand, to elaborate time-dependent or non-BPS solutions, we 
need to study second derivative equation of motion.
The equation of motion for $Z^A$ with $W_A = A_{\mu} = \hat{A}_{\mu} = 0$ is 
obtained as 
\begin{eqnarray}
\Box Z^A &=& \frac{4 \pi^2}{k^2} 
\left\{
3 (Z^B Z^{\dagger}_B)^2 Z^A + 3 Z^A (Z^{\dagger}_B Z^B)^2 - 2 Z^A Z^{\dagger}_B
(Z^C Z^{\dagger}_C) Z^B \right. \nonumber \\
& & \left.
 \qquad \qquad - 2 (Z^B Z^{\dagger}_B) Z^A (Z^{\dagger}_C Z^C) - 2 Z^B (Z^{\dagger}_C 
Z^C) Z^{\dagger}_B Z^A 
\right\}, \label{eom}
\end{eqnarray}
where $\Box = \partial_{\mu} \partial^{\mu}$.
Note that this equation is Lorentz invariant. 
We can easily show that the BPS equation (\ref{massless_BPS}) is consistent with this equation of motion.
Let us analyze a solution of the equation of motion with specific ansatz.
As in the case of the BPS configurations, consider an ansatz
\begin{eqnarray}
Z^A = f (x) S^A, \quad f (x) \in \mathbf{C} \label{ansatz1}
\end{eqnarray}
with the BPS matrices $S^A$,
then due to the property of the BPS matrices, the 
right hand side of the equation of motion becomes proportional to $S^A$ 
(see appendix) and the equation (\ref{eom}) reduces to the 
equation for $f (x)$,
\begin{eqnarray}
\Box f (x) = \frac{12\pi^2}{k^2} f (x) |f(x)|^4.
\end{eqnarray}

On the other hand, it is possible to construct one parameter deformation 
of the ABJM model preserving an $SU(2) \times SU(2) \times U(1)_R$ 
global symmetry and $\mathcal{N} = 6$ supersymmetry. 
It is just the mass deformed ABJM model \cite{HoLeLeLePa, GoRoRaVe}.
The equation of motion for the mass deformed ABJM model is 
\begin{eqnarray}
\Box Z^A &=& \frac{4 \pi^2}{k^2} 
\left\{
3 (Z^B Z^{\dagger}_B)^2 Z^A + 3 Z^A (Z^{\dagger}_B Z^B)^2 - 2 Z^A Z^{\dagger}_B
(Z^C Z^{\dagger}_C) Z^B \right. \nonumber \\
& & \left.
 \qquad \qquad - 2 (Z^B Z^{\dagger}_B) Z^A (Z^{\dagger}_B Z^B) - 2 Z^B (Z^{\dagger}_C 
Z^C) Z^{\dagger}_B Z^A 
\right\} \nonumber \\
& & - \frac{8 \pi m}{k} \left\{
Z^B Z^{\dagger}_B Z^A - Z^A Z^{\dagger}_B Z^B 
\right\} + m^2 Z^A, \label{massive_eom}
\end{eqnarray}
where $m$ is the mass of $Z^A$. 
Here, we have dropped the gauge fields and $W_A$.
Plugging the ansatz (\ref{ansatz1}) back into the equation 
(\ref{massive_eom}), we find the equation for the massive case,
\begin{eqnarray}
\Box f (x) = \frac{12\pi^2}{k^2} f |f|^4 (x) - \frac{8 \pi m}{k} f |f|^2 (x) 
+ m^2 f (x). \label{m_eom}
\end{eqnarray}
As in the massless case, the BPS matrices are linearized on both sides 
of the equation (\ref{massive_eom}). 
On the other hand, the 1/2 BPS equation of the massive ABJM model is \cite{HaLi}\begin{eqnarray}
\partial_2 Z^A = - \frac{2 \pi}{k} \left\{
(Z^B Z^{\dagger}_B) Z^A - Z^A (Z^{\dagger}_B Z^B)
\right\} + m Z^A.
\end{eqnarray}
This is consistent with the massive equation of motion (\ref{massive_eom})
. If we assume that $f (x_2)$ is a real function, the BPS solutions of the equation (\ref{m_eom}) are given by
\begin{eqnarray}
f_{\pm} (x_2) = \sqrt{\frac{k m}{2 \pi}} 
\left(
1 \pm e^{-2 m x_2}
\right)^{- \frac{1}{2}},
\label{massive_BPS}
\end{eqnarray}
where $f_+$ is a domain wall interpolating a trivial vacuum and 
a non-trivial fuzzy $S^3$ vacuum while $f_{-}$ is a deformed fuzzy funnel \cite{HaLi}.
The ansatz (\ref{ansatz1}) preserves $SU(2)$ symmetry. 

Note that any scalar fields satisfying the equations (\ref{eom}), (\ref{massive_eom})
should satisfy the ``Gauss' law'' constraints,
\begin{eqnarray}
& & (\partial_{\mu} Z^A) Z^{\dagger}_A - Z^A (\partial_{\mu} Z^A)^{\dagger} 
= 0, \nonumber \\
& & (\partial_{\mu} Z^A)^{\dagger} Z^A 
 - Z^{\dagger}_A (\partial_{\mu} Z^A) = 0.
\end{eqnarray}
These constraints are automatically satisfied for the ansatz 
(\ref{ansatz1}) provided 
\begin{eqnarray}
f (x) = e^{i \phi} g (x), \quad g (x) \in \mathbf{R},
\end{eqnarray}
with constant phase $\phi \in \mathbf{R}$.

A comment is in order. Because the equation of motions for massless and 
massive cases are manifestly Lorentz invariant, we can construct 
Lorentz boosted solutions from known static BPS solutions.
A few examples of these boosted solutions are presented in the appendix B.

\section{Oscillating fuzzy $S^3$}
In this section, we consider time evolutions of fuzzy $S^3$ 
in the ABJM model by solving the equation of motion with the 
ansatz of purely time-dependent function $f (t)$.
Several works on time evolutions of fuzzy spheres have been studied in the context of D-brane effective theories. 
In particular, collapsing fuzzy $S^2$ \cite{PaRaTo, PaRa, McPaRaSp} 
and $S^3, S^5$ \cite{PaRa2} solutions were analyzed by 
the Dirac-Born-Infeld (DBI) equation of motion in the fully non-linear level.
We will show that a similar collapsing fuzzy sphere exists in the 
massless ABJM model. Moreover, in the massive ABJM model, there are 
variety of solutions depending on the value of the mass parameter.

\subsection{Massless case}
\begin{figure}[t]
\begin{center}
\includegraphics[scale=.7]{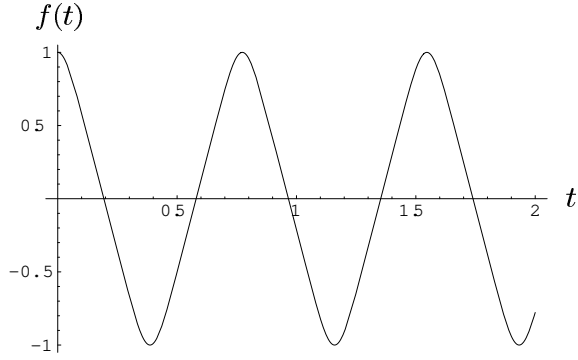}
\end{center}
\caption{A numerical plot of the solution for the equation (\ref{time-fuzzy}) with $k =1,\, f(t=0) = f_0 = 1,\, f' (t=0) = 0$.
}
\label{t-fuzzy_numerical}
\end{figure}

Consider an ansatz $Z^A = f(t) S^A$ with real $f (t)$. The equation 
(\ref{eom_scalar}) reduces to
\begin{eqnarray}
- \partial_t^2 f (t) = 3 \alpha f^5 (t),
\label{time-fuzzy}
\end{eqnarray}
where $\alpha = \frac{4 \pi^2}{k^2}$.
This can be rewritten as 
\begin{eqnarray}
& & \dot{f}^2 (t) = - \alpha f^6 (t) + c_0, \label{t-fuzzy}
\end{eqnarray}
where $c_0$ is an integration constant and the dot stands for the time derivative.
Assuming an initial condition $\dot{f} (t=t_0) = 0, f (t=t_0) = f_0 > 0$, we have $c_0 = \alpha f_0^6$. The solution is valid only in the region $0< |f| < f_0$. 
The physical radius of the fuzzy $S^3$ is 
\begin{eqnarray}
R^2 = \frac{2}{N} \mathrm{Tr} \left[X^{\dagger}_A X^A \right] 
= \frac{2}{N T_2} \mathrm{Tr} \left[S^{\dagger}_A S^A \right] f^2 (t)
= \frac{2 (N-1)}{T_2} f^2 (t).
\end{eqnarray}
Here we have used the relation $\mathrm{Tr} [S^{\dagger}_A S^A] = N 
(N-1)$ and physical coordinate $X^A \equiv \sqrt{\frac{1}{T_2}} Z^A$ \cite{HaLi}. The constant $T_2$ is the tension of an M2-brane.
A numerical plot of the solution to the equation (\ref{time-fuzzy}) is illustrated in fig.\,\ref{t-fuzzy_numerical}.
This is an oscillating fuzzy sphere 
and we call this type of fuzzy sphere ``collapsing'' fuzzy sphere 
since it collapses to zero size then expands again.
A similar analysis of collapsing fuzzy $S^2$ from the viewpoint of the DBI
equation of motion in D2-brane world-volume was carried out in \cite{PaRa} 
where the time-dependent solution is described by Jacobi elliptic 
functions exhibiting a large-small duality of the fuzzy sphere radius.

Let us evaluate the decay time of the fuzzy sphere by analytically 
solving the equation (\ref{t-fuzzy}). This can be done by 
separating $f>0$ and $f<0$ cases. 
For $f > 0$ case, the equation (\ref{t-fuzzy}) is integrated to give
\begin{eqnarray}
\frac{1}{\sqrt{\alpha}} \int^f_{f_0} \frac{d f}{\sqrt{f^6_0 - f^6}} 
= \pm \int^t_{t_0} d t. 
\label{massless_eq_integrated}
\end{eqnarray}
The left hand side of the above equation is evaluated as
\begin{eqnarray}
\frac{1}{6 f_0^2 \sqrt{\alpha}} \left[ B \left(\frac{f^6}{f^6_0}, \frac{1}{6}, \frac{1}{2} \right)
- B \left(\frac{1}{6}, \frac{1}{2} \right)
\right],
\quad f^6/f^6_0 \le 1.
\end{eqnarray}
Here we have used the definition of (incomplete) beta function
\begin{eqnarray}
& & B (a,b) = \int^1_0 \! dt \ t^{a-1} (1 -t)^{b-1} 
= \int^{\infty}_0 \! dt \ t^{a-1} (t+1)^{-a-b}
= \int^{\infty}_1 \! dt \ t^{-a-b} (t-1)^{a-1}, \nonumber \\
& & B (z,a,b) = \int^z_0 \! dt \ t^{a-1} (1 -t)^{b-1}
 = \int^{\frac{z}{1-z}}_0 \! dt \ t^{a-1} (t+1)^{-a-b}
 = \int^{\frac{1}{1-z}}_1 \! dt \ t^{a-1} (t-1)^{-a-b}
, \nonumber \\
& &  0< \mathrm{Re} (z) < 1, \ a,b >0.
\end{eqnarray}
Thus the equation (\ref{massless_eq_integrated}) is
\begin{eqnarray}
\frac{1}{6 f_0^2 \sqrt{\alpha}} 
B \left(\frac{1}{6}, \frac{1}{2} \right)
\left\{
I \left( \frac{f^6}{f^6_0}, \frac{1}{6}, \frac{1}{2} \right) - 1 
\right\}
= \pm (t - t_0).
\end{eqnarray}
Here $I(z,a,b)$ is the regularized beta function defined by
\begin{eqnarray}
I (z, a, b) = \frac{B (z,a,b)}{B (a,b)}.
\end{eqnarray}
From this expression, we find the following 
analytic solution in the region $f \ge 0$\footnote{The full global solution can be obtained by taking the 
massless limit of the solution (\ref{sol_int}).}
\begin{eqnarray}
f (t) = f_0 \left[
I^{-1} \left(\tau, \frac{1}{6}, \frac{1}{2} \right)
\right]^{\frac{1}{6}}, \quad 
\tau = \pm \frac{12 \pi f_0^2}{k} \frac{t-t_0}{B \left(\frac{1}{6}, 
\frac{1}{2} \right)} + 1, 
\label{sol1}
\end{eqnarray}
where $I^{-1} (z, a, b)$ is the inverse regularized beta function.
The solution for $f < 0$ region can be obtained by reversing the sign 
in front of $f_0$ in the equation (\ref{sol1}) and changing $t 
\leftrightarrow -t$.
The decay time $T = t_d - t_0, \ f (t_d) = 0$ is evaluated as 
\begin{eqnarray}
& & T =
\frac{1}{6 f^2_0 \sqrt{\alpha}} 
\left| B \left(0, \frac{1}{6}, \frac{1}{2} \right)
- B \left(\frac{1}{6}, \frac{1}{2} \right) \right| 
= 
\frac{k}{12 \pi f^2_0} 
\frac{\Gamma \left(\frac{1}{6} 
\right)
\Gamma \left(\frac{1}{2}\right)}{\Gamma \left( \frac{2}{3} \right)}, \nonumber \\
& &  
\frac{\Gamma \left(\frac{1}{6} 
\right)
\Gamma \left(\frac{1}{2}\right)}{\Gamma \left( \frac{2}{3} \right)} 
\sim 7.28595.
\end{eqnarray}
A conserved energy of this oscillating fuzzy $S^3$ is also evaluated analytically as 
\begin{eqnarray}
E = \int d^3 x \mathrm{Tr} 
[
S^{\dagger}_A S^A 
]
\left(
(\partial_t f)^2 + \frac{4 \pi^2}{k^2} f^6 
\right) 
= \frac{4 \pi^2}{k^2} f_0^6 N (N-1) \int d^3 x 
\end{eqnarray}
For $N = 1$, the energy vanishes indicating the fact 
that this solution is purely due to the non-abelian nature of 
multiple M2-branes.

\subsection{Massive case}
Once we introduce a mass term for $Z^A$, the 
ansatz $Z^A = f (t) S^A$ satisfies the equation 
\begin{eqnarray}
- \partial^2_t f (t) = \frac{12 \pi^2}{k^2} f^5 (t) - \frac{8 \pi m}{k} f^3 (t)
+ m^2 f(t), \quad f (t) \in \mathbf{R}. \label{time_massive}
\end{eqnarray}
Assuming an initial condition $\dot{f} (t_0) = 0, f (t_0) = f_0$, 
the equation (\ref{time_massive}) reduces to 
\begin{eqnarray}
\dot{f}^2 &=& - \alpha (f^6 - f^6_0) + \beta (f^4 - f^4_0) - \gamma (f^2 - 
f^2_0), \label{massive_eom_initial}
\end{eqnarray}
where we have defined
\begin{eqnarray}
& & \alpha = \frac{4 \pi^2}{k^2}, \ \beta = \frac{4 \pi m}{k}, \ \gamma = m^2.
\end{eqnarray}
The equation of motion can be rewritten as 
\begin{eqnarray}
\dot{\tau}^2 &=& - 4 \alpha \tau (\tau - f_0^2 ) 
\left[
\tau^2 + \left(f_0^2 - \frac{\beta}{\alpha} \right) \tau
+ \frac{1}{\alpha} (\alpha f^4_0 - \beta f^2_0 + \gamma)
\right], \label{massive_osc_eq}
\end{eqnarray}
where we have defined $\tau = f^2$.
We first consider $f > 0$ region. 
In this case, the equation (\ref{massive_osc_eq}) becomes
\begin{eqnarray}
\int^{f^2}_{f_0^2} \! \frac{d \tau}{\sqrt{- \tau (\tau - f_0^2) 
\left\{
\tau^2 + \left(f_0^2 - \frac{\beta}{\alpha} \right) \tau 
+ \frac{1}{\alpha} 
(\alpha f^4_0 - \beta f^2_0 + \gamma)
\right\}}}
= \pm 2 \sqrt{\alpha} \int^t_{t_0} \! d t.
\label{integrand}
\end{eqnarray}
The integral on the left hand side of the equation (\ref{integrand}) is 
separately evaluated for the following two cases,
(A) $\tau^2 + \left(f_0^2 - \frac{\beta}{\alpha} \right) \tau + 
\frac{1}{\alpha} \left( \alpha f_0^4 - \beta f_0^2 + \gamma \right) = 0$ 
has real solutions, (B) $\tau^2 + \left(f_0^2 - \frac{\beta}{\alpha} \right) \tau + 
\frac{1}{\alpha} \left( \alpha f_0^4 - \beta f_0^2 + \gamma \right) = 0$ 
has no real solutions. 
For the case (A), $f_0$ should satisfy 
$0 < f_0^2 \le \frac{2 \beta}{3 \alpha}$ while 
$f_0^2 > \frac{2 \beta}{3 \alpha}$ for the case (B). 
If $f_0 = 0$, there is only a trivial solution $f = 0$.

\subsubsection{Case (A)}
\begin{figure}[t]
\begin{center}
\includegraphics[scale=.7]{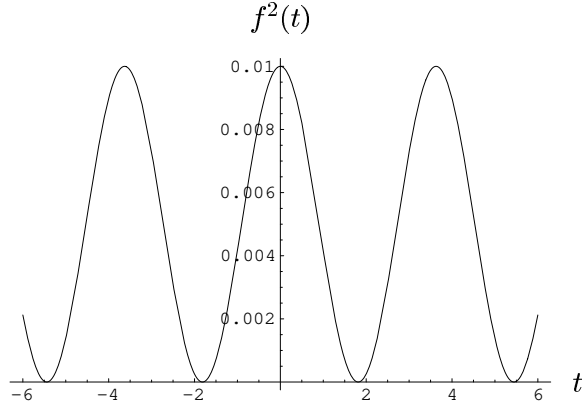}
\end{center}
\caption{An analytic plot of the solution (\ref{massive_ansol1}). $k = m 
= 1$, $f_0 = 0.1$, $t_0 = 0$. Plus sign in $\pm$ is chosen.} 
\label{massive_ansol1_plot}
\end{figure}

\begin{figure}[t]
\begin{center}
\includegraphics[scale=.7]{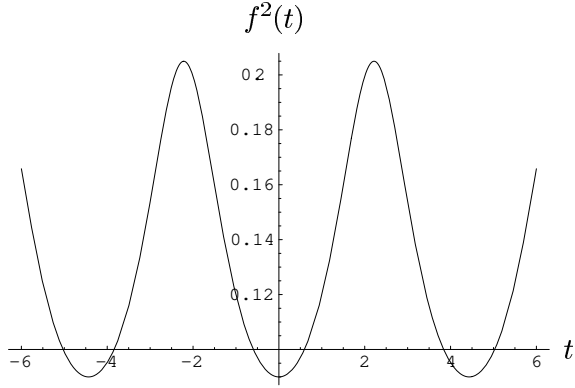}
\end{center}
\caption{An analytic plot of the solution (\ref{massive_ansol2}) with $k = m 
= 1$, $f_0 = 0.3$, $t_0 = 0$, $g_1 = 0.205$, $g_2 = 0.0233$. The plus sign in $\pm \sqrt{\alpha}$ is chosen.} 
\label{massive_ansol2_plot}
\end{figure}

\begin{figure}[t]
\begin{center}
\includegraphics[scale=.7]{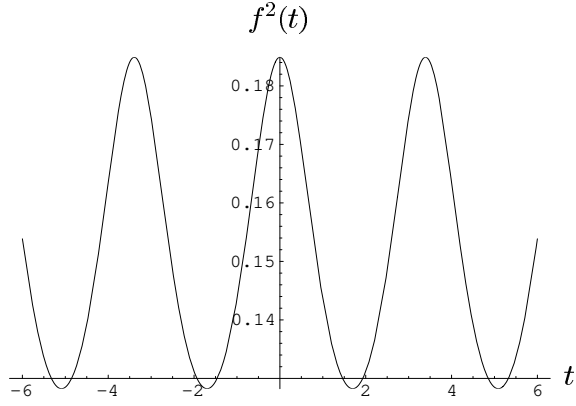}
\end{center}
\caption{An analytic plot of the solution (\ref{massive_ansol3}) with $k = m 
= 1$, $f_0 = 0.43$, $t_0 = 0$, $g_1 = 0.128$, $g_2 = 0.00517$. The plus sign in $\pm \sqrt{\alpha}$ is chosen.} 
\label{massive_ansol3_plot}
\end{figure}

\begin{figure}[t]
\begin{center}
\includegraphics[scale=.7]{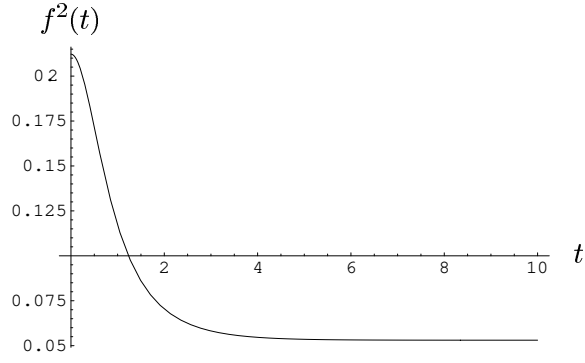}
\end{center}
\caption{An analytic profile for the solution (\ref{sol-vi}) with
$k=m=1$, $f_0=\sqrt{\frac{2}{3\pi}}$, $t_0 = 0$.} 
\label{sol_damp}
\end{figure}

In this situation, the integrand on the left hand side of the 
equation (\ref{integrand}) can be written as the following form
\begin{eqnarray}
\frac{1}{\sqrt{- \tau (\tau - f_0^2) (\tau - g_1) 
(\tau - g_2)}}, 
\end{eqnarray}
Here, the constants $g_1, g_2$ are given by
\begin{eqnarray}
g_1 &=& \frac{1}{2} \left[
\left(- f_0^2 + \frac{\beta}{\alpha} \right)
+ f_0 \sqrt{- 3 f^2_0 + \frac{2 \beta}{\alpha}}
\right], \\
g_2 &=& \frac{1}{2} \left[
\left(- f_0^2 + \frac{\beta}{\alpha} \right)
- f_0 \sqrt{- 3 f^2_0 + \frac{2 \beta}{\alpha}}
\right].
\end{eqnarray}
The existence of the real roots $g_1, g_2$ is guaranteed by the constraint
\begin{eqnarray}
0 < f^2_0 \le \frac{2 \beta}{3\alpha}.
\end{eqnarray}
In this region of $f_0^2$, both $g_1$ and $g_2$ are positive semi-definite.
The behavior of the solution is different depending on the values of $g_1$ and $g_2$. 
In this region, there are following possibilities according to the value of 
$f_0^2$,
\begin{eqnarray}
& & \textrm{(i)} \quad g_1 > g_2 > f_0^2 > 0, 
\ \left(
0 < f_0^2 < \frac{1}{6} \frac{\beta}{\alpha}
\right), \nonumber \\
& & \textrm{(ii)} \quad 
g_1 > g_2 = f_0^2 > 0, \
\left(
f_0^2 = \frac{1}{6} \frac{\beta}{\alpha}
\right), 
\nonumber \\
& & \textrm{(iii)} \quad g_1 > f_0^2 > g_2 > 0, 
\ \left(
\frac{1}{6} \frac{\beta}{\alpha} < f_0^2 < \frac{1}{2} \frac{\beta}{\alpha}
\right), \nonumber \\
& & \textrm{(iv)} \quad 
g_1 = f_0^2 > g_2 > 0, \
\left(
f_0^2 = \frac{1}{2} \frac{\beta}{\alpha}
\right), 
\nonumber \\
& & \textrm{(v)} \quad f_0^2 > g_1 > g_2 > 0, 
\ 
\left(
\frac{1}{2} \frac{\beta}{\alpha} < f_0^2 < \frac{2}{3} 
\frac{\beta}{\alpha}
\right), \nonumber \\
& & \textrm{(vi)} \quad 
f_0^2 > g_1 = g_2 > 0, \
\left(
f_0^2 = \frac{2}{3} \frac{\beta}{\alpha}
\right). \nonumber 
\end{eqnarray}
Solutions for the cases (i)-(vi) are studied separately in below.
\\
\\
\underline{\textbullet \ Case (i) $\qquad g_1 > g_2 > f_0^2 > 0$}
\\
\\
For the case (i), the integration is evaluated as
\begin{eqnarray}
- \int^{f_0^2}_{f^2} \frac{d \tau}{\sqrt{-(\tau - g_1) (\tau - g_2) 
(\tau - f_0^2) \tau}} 
= - \frac{2}{\sqrt{(g_1 - f_0^2) g_2}} 
F \left(
\mathrm{arcsin} \sqrt{\frac{g_2}{f_0^2} \frac{f_0^2 - f^2}{g_2 - f^2}}, \kappa_1
\right), \label{mass_int_1}
\end{eqnarray}
where
\begin{eqnarray}
f_0^2 > f^2 > 0, \quad \kappa_1 = \sqrt{\frac{g_1 - g_2}{g_1 - f_0^2} \frac{f_0^2}{g_2}}.
\end{eqnarray}
Here $F (\varphi, \kappa)$ is an elliptic integral with modulus $\kappa$ defined by 
\begin{eqnarray}
F (\varphi, \kappa) = \int^{\varphi}_0 \frac{d \theta}{\sqrt{1 - \kappa^2 \sin^2 \theta}},
\quad F (\mathrm{arcsin} (x), \kappa) = \mathrm{sn}^{-1} x.
\label{massive_integral1}
\end{eqnarray}
Here $\mathrm{sn} (x)$ is the Jacobi's elliptic function.
From the expression (\ref{mass_int_1}), we obtain
\begin{eqnarray}
f^2 (t) = g_2 f_0^2
\frac{ \mathrm{sn}^2 \left[ \mp \sqrt{\alpha} \sqrt{g_2 (g_1 - f_0^2) }
(t-t_0) \right] - 1 }{f_0^2 \mathrm{sn}^2 \left[ \mp \sqrt{\alpha} \sqrt{g_2 (g_1 - f_0^2) 
}(t-t_0) \right] - g_2}.
\label{massive_ansol1}
\end{eqnarray}
The solution is the positive root of the equation (\ref{massive_ansol1}).
For $f < 0$, the solution is given by the square root of (\ref{massive_ansol1}) with minus sign and replacement $t \leftrightarrow 
-t$. 
An analytic profile of the solution can be found in fig.\,\ref{massive_ansol1_plot}.
This solution represents a collapsing fuzzy sphere with the decay time 
$T$ given by
\begin{eqnarray}
T = \frac{1}{\sqrt{\alpha} \sqrt{g_1 - f_0^2} g_2}
F \left( \frac{\pi}{2}, \kappa_1
\right).
\end{eqnarray}
\\
\\
\underline{\textbullet \ Case (ii) $\qquad g_1 > g_2 = f_0^2 > 0,
  f_0^2 = \frac{1}{6} \frac{\beta}{\alpha}$}
\\
\\
For the case (ii), the equation of motion (\ref{massive_eom_initial}) is rewritten as 
\begin{eqnarray}
\dot{f}^2 = \frac{1}{108 \alpha} 
(6 \alpha f^2 - \beta)^2 (3 \alpha f^2 - 2 \beta).
\end{eqnarray}
Because $3 \alpha f^2 - 2 \beta < 3 g_1 - 2 \beta = 0$ at $f_0^2 = 
\frac{1}{6} \frac{\beta}{\alpha}$, the fact that
the right hand side of the above equation should be positive 
semi-definite requires that the solution should be $f^2 = f_0^2 = \frac{1}{6} \frac{\alpha}{\beta}$.
Although this is an extremum point of the potential, this point is a local maximum and the configuration is unstable.
\\
\\
\underline{\textbullet \ Case (iii) $\qquad g_1 > f_0^2 > g_2 > 0$}
\\
\\
In the case (iii), the integration is evaluated as 
\begin{eqnarray}
\int^{f^2}_{f_0^2} \frac{d \tau}{\sqrt{-(\tau - g_1) (\tau - f_0^2) 
(\tau - g_2)  \tau}} 
= \frac{2}{\sqrt{(g_1 - g_2) f_0^2}} 
F \left( 
\mathrm{arcsin} \sqrt{\frac{g_1 - g_2}{g_1 - f_0^2} \frac{f^2 - 
f_0^2}{f^2 - g_2}}, \kappa_2
\right),
\end{eqnarray}
where
\begin{eqnarray}
g_1 > f^2 > f_0^2, \quad \kappa_2 = \sqrt{\frac{g_1 - f_0^2}{g_1 - g_2} 
\frac{g_2}{f_0^2}}.
\end{eqnarray}
The solution is 
\begin{eqnarray}
f^2 (t) = 
\frac{g_2 (g_1 - f_0^2) \mathrm{sn}^2 \left[ \pm \sqrt{\alpha} f_0 
\sqrt{g_1-g_2} (t-t_0) \right] - f_0^2 (g_1 - g_2)}{(g_1 - f_0^2) 
\mathrm{sn}^2 \left[ \pm \sqrt{\alpha} f_0 
\sqrt{g_1-g_2} (t-t_0) \right]  - (g_1 - g_2)}.
\label{massive_ansol2}
\end{eqnarray}
The solution for $f < 0$ region can be obtained as well as for the case (i). 
Note that when the numerator in (\ref{massive_ansol2}) 
goes to zero, the denominator also goes to zero and $f^2(t)$ remains 
finite value. This means that the solution does not collapse into zero size 
as can be seen in fig.\,\ref{massive_ansol2_plot}. 
$f^2$ oscillates between $g_1$ and $f_0^2$ meaning oscillating fuzzy 
$S^3$ with finite radius.
\\
\\
\underline{\textbullet \ Case (iv) $\qquad g_1 = f_0^2 > g_2 > 0,
  f_0^2 = \frac{1}{2} \frac{\beta}{\alpha}$ }
\\
\\
In this case, the right hand side of the 
equation (\ref{massive_eom_initial}) is positive semi-definite only at the 
point $f^2 = f_0^2 = \frac{1}{2} \frac{\beta}{\alpha}$. 
This is just the vacuum configuration found in \cite{GoRoRaVe}.
\\
\\
\underline{\textbullet \ Case (v) $\qquad f_0^2 > g_1 > g_2 > 0,
 \frac{1}{2} \frac{\beta}{\alpha} < f_0^2 < \frac{2}{3} \frac{\beta}{\alpha} $}
\\
\\
In this case, the integration is evaluated as
\begin{eqnarray}
- \int^{f_0^2}_{f^2} 
\frac{d \tau}{\sqrt{- (\tau - f_0^2) (\tau - g_1) 
(\tau - g_2)  \tau}} 
= - \frac{2}{\sqrt{(f_0^2 - g_2) g_1}}
F 
\left(
\mathrm{arcsin} \sqrt{\frac{g_1}{f_0^2 - g_1} \frac{f_0^2 - f^2}{f^2}}
, \kappa_3
\right),
\end{eqnarray}
where 
\begin{eqnarray}
f_0^2 > f^2 > g_1, \quad 
\kappa_3 = \sqrt{\frac{f_0^2 - g_1}{f_0^2 - g_2} \frac{g_2}{g_1}}.
\end{eqnarray}
The solution is 
\begin{eqnarray}
f^2 (t) = \frac{g_1 f_0^2}{(f_0^2 - g_1) \mathrm{sn}^2 \left[ \mp 
\sqrt{\alpha} \sqrt{(f_0^2 - g_2) g_1} (t-t_0) \right] + g_1}.
\label{massive_ansol3}
\end{eqnarray}
An analytic profile can be found in fig.\,\ref{massive_ansol3_plot}.
Note that because the numerator in the right hand side of the equation (\ref{massive_ansol3}) 
is always positive, this solution does not collapse into zero size but oscillates within non-zero values of the radius. 
\\
\\
\underline{\textbullet \ Case (vi) $ \qquad f_0^2 > g_1 = g_2 > 0,
  f_0^2 = \frac{2}{3} \frac{\beta}{\alpha} $}
\\
\\
In this case $g_1 = g_2 = \frac{1}{6} \frac{\beta}{\alpha} = \frac{1}{4} 
f_0^2$.
The integration is 
\begin{eqnarray}
\int^{f^2}_{f_0^2} \frac{d \tau}{|\tau - g_1| 
\sqrt{ - \tau (\tau - f_0^2)}}.
\end{eqnarray}
Consider $\tau > g_1$ case. Then the integration is evaluated as 
\begin{eqnarray}
\left. \frac{1}{\sqrt{\kappa}} 
\log \left|
\frac{1}{\tau - g_1} 
\left(
\frac{1}{2} f_0^2 (\tau - g_1) + \frac{3}{8} f_0^4 
- 2 \sqrt{ 
\frac{3}{16} f_0^4 \left( - \tau (\tau - f_0^2)\right)
}
\right)
\right|
\right|^{f^2}_{f_0^2},
\end{eqnarray}
where $\kappa = \frac{3}{16} f_0^4$.
From this expression, we have the following solution
\begin{eqnarray}
f^2 (t) = \frac{f_0^2}{4} 
\frac{(1 + \exp \left[ \pm 2 \sqrt{\alpha \kappa} (t-t_0) \right] 
)^2}{1 - \exp \left[ \pm 2 \sqrt{\alpha \kappa} (t-t_0) \right] 
+ \exp \left[ \pm 4 \sqrt{\alpha \kappa} (t-t_0) \right]}, \quad (t \ge t_0).
\label{sol-vi}
\end{eqnarray}
This solution is neither oscillating nor collapsing but 
shrinking down to a finite radius at $t \to \infty$. Indeed, at large 
$t$, we have
\begin{eqnarray}
\lim_{t \to \infty} f^2 (t) = \frac{f_0^2}{4}.
\end{eqnarray}
An analytic profile can be found in fig.\,\ref{sol_damp}.
\subsubsection{Case (B)}
In this case, the integral on the left hand side in the equation 
(\ref{integrand}) can be rewritten as 
\begin{eqnarray}
\int^{f^2}_{f_0^2} \frac{d \tau}{\sqrt{\varphi (\tau)}} 
= \int^{(B-f^2)/(A-f^2)}_{(B-f_0^2)/(A-f_0^2)} 
\frac{A-B}{\sqrt{|\varphi (A)|}} \frac{d t}{\sqrt{ (t^2 - \mu^2) (t^2 + 
\nu^2)}},
\label{integral_nosol}
\end{eqnarray}
where we have defined 
\begin{eqnarray}
\varphi (x) &=& F (x) G (x), \nonumber \\
F (x) &=& - x^2 + f_0^2 x, \nonumber \\
G (x) &=& \frac{1}{\alpha} 
\left(
x^2 + \left(
f_0^2 - \frac{\beta}{\alpha} 
\right) x
+ \frac{1}{\alpha} 
\left(
\alpha f_0^4 - \beta f_0^2 + \gamma
\right)
\right),
\end{eqnarray}
and 
\begin{eqnarray}
A &=& \frac{1}{\beta - 2 \alpha f_0^2} 
\left[
j (f_0^2) - \sqrt{h (f_0^2)}
\right], \\
B &=& \frac{1}{\beta - 2 \alpha f_0^2} 
\left[
j (f_0^2) + \sqrt{h (f_0^2)}
\right].
\end{eqnarray}
The functions $ j (x), h (x)$ are given by 
\begin{eqnarray}
j (x) &\equiv& 
\alpha x^2 - \beta x + \gamma, \\
h (x) &\equiv& 
3 \alpha^2 x^4 - 5 \alpha \beta x^3 + 4 \alpha \gamma x^2 
+ 2 \beta^2 x^2 - 3 \beta \gamma x + \gamma^2.
\end{eqnarray}
\begin{eqnarray}
\mu^2 = - \frac{F(B)}{F(A)} > 0, \quad \nu^2 = \frac{G(B)}{G(A)} > 0.
\end{eqnarray}
Note that because $\beta^2 - 4 \alpha \gamma = 0$, $j (x)$ is positive 
definite and due to the condition $f_0^2 > \frac{\beta}{2\alpha}$, we 
have $\beta - 2 \alpha f_0^2 < 0$. From these facts, we can see 
$A > 0, B < 0$ and $G (A) > 0, \quad G (B) > 0, \quad F (B) < 0$.
Although, there are two possibilities 
\begin{eqnarray}
& & F (A) > 0, \quad (0 < A < f_0^2), \\ 
& & F (A) < 0, \quad (A > f_0^2),
\end{eqnarray}
we consider $F (A) > 0$ case, otherwise the integral becomes complex valued.
The integral (\ref{integral_nosol}) can be rewritten by the following 
change of variable,
\begin{eqnarray}
t^2 = \mu^2 (1 - u^2).
\end{eqnarray}
Then the integral (\ref{integral_nosol}) is
\begin{eqnarray}
\int^{\lambda}_{\lambda_0} \frac{A-B}{\sqrt{|\varphi (A) |}} 
\frac{c du}{\sqrt{(1 - u^2) (1 - \kappa_4^2 u^2)}} 
= c \frac{A-B}{\sqrt{|\varphi (A)|}}
\left[ \frac{}{}
F (\mathrm{arcsin} (\lambda), \kappa_4) - F (\mathrm{arcsin} (\lambda_0), \kappa_4)
\right]. \label{sol_int}
\end{eqnarray}
Here
\begin{eqnarray}
& & \kappa_4^2 = \frac{\mu^2}{\mu^2 + \nu^2}, \quad 
c = \pm \frac{2 \mu u}{\sqrt{\mu^2 + \nu^2} |\mu| |u|}, \\
& & \left(
\frac{B-f^2}{A-f^2}
\right)^2 
= \mu^2 (1 - \lambda^2), \quad 
\left(
\frac{B-f_0^2}{A-f_0^2}
\right)^2 
= \mu^2 (1 - \lambda_0^2).
\end{eqnarray}
Note that $\pm$ in the expression of $c$ is determined 
by whether $t$ is positive ($+$) or negative ($-$).
Finding the explicit form of the solution $f (t)$ from the equation (\ref{sol_int}) is straightforward but we do not present it here.
However, it is obvious that this solution is an oscillating fuzzy sphere 
which can be collapsed into zero size as we will see below.
Summary of these solutions is found in table \ref{summary_massive_solution}.
\begin{table}[t]
\begin{center}
\begin{tabular}{|l||l|l|}
\hline
value of $f_0^2$ & Solution & Remarks
 \\
\hline \hline
$f_0^2 = 0$ & $f^2 = 0$ & Vacuum \\
\hline
$0 < f_0^2 < \frac{1}{6} \frac{\beta}{\alpha}$ & Eq.~(\ref{massive_ansol1}) & Collapsing fuzzy $S^3$ \\
\hline
$f_0^2 = \frac{1}{6} \frac{\beta}{\alpha}$ & $f^2 = \frac{1}{6} 
\frac{\beta}{\alpha}$ & unstable point \\
\hline
$\frac{1}{6} < f_0^2 < \frac{1}{2} \frac{\beta}{\alpha}$ & Eq.~(\ref{massive_ansol2}) & 
Oscillating fuzzy $S^3$ with finite radius \\
\hline
$f_0^2 = \frac{1}{2} \frac{\beta}{\alpha}$ & $f^2 = \frac{1}{2} 
\frac{\beta}{\alpha}$ & Vacuum \\
\hline
$\frac{1}{2} \frac{\beta}{\alpha} < f_0^2 < \frac{2}{3} 
\frac{\beta}{\alpha}$ & Eq.~(\ref{massive_ansol3}) & Oscillating fuzzy $S^3$ with finite radius 
\\
\hline
$f_0^2 = \frac{2}{3} \frac{\beta}{\alpha}$ & Eq.~(\ref{sol-vi})  & Shrinking fuzzy $S^3$ \\
\hline
$f_0^2 > \frac{2}{3} \frac{\beta}{\alpha}$ & Eq.~(\ref{sol_int}) & Collapsing fuzzy $S^3$ 
\\
\hline
\end{tabular}
\caption{Solutions for the massive case}
\label{summary_massive_solution}
\end{center}
\end{table}

\begin{figure}[t]
\begin{center}
\includegraphics[scale=.7]{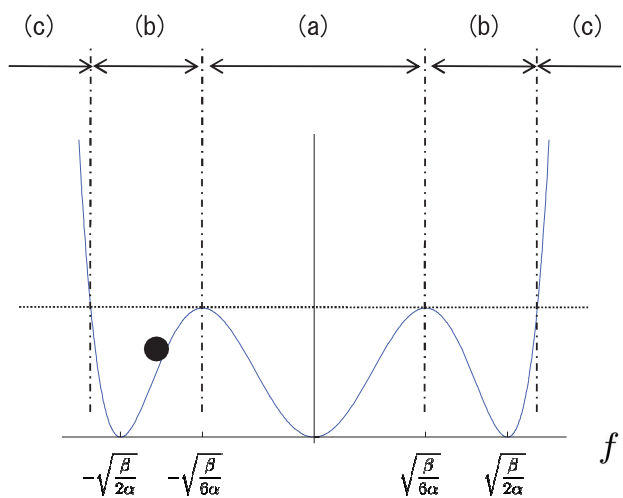}
\end{center}
\caption{The behavior of the equation of motion is able to
be understood as the behavior of particle obeying classical equation of motion 
with the 6th-order potential.} 
\label{motion_in_6th-order_potential}
\end{figure}

Because the nature of equations of motion comes from the 
6th-order polynomial function for the scalar potential,
the above characteristics of the solutions is inherently
understood as the motion of a particle which obey classical
equation of motion in the 6th-order polynomial potential.
Setting off a particle with zero-velocity from a certain point
that is given by $f_0$, on the one dimensional potential curve, 
its trajectory shows the behavior of the corresponding solution
(fig.\,\ref{motion_in_6th-order_potential}).
Since the potential blows up at infinity, it must be bounded.
If the initial point is within the region (a),
namely $|f_0|< \sqrt{\beta /2 \alpha}$,
the particle begins to oscillate around the origin as in the case (i),
except for the origin $f_0=0$.
Then it passes across the origin in its cycle,
the radius of the fuzzy $S^3$ collapses.
In contrast, if the initial point is within the region (b),
the particle oscillates around 
one of the local minima at $f_0 = \pm \sqrt{\beta /2 \alpha}$,
and thus the radius never collapses in this case, that is responsible for
the case (iii) or (v).
Within the region (c), the amplitude of the oscillating motion is large enough
to get over both local maxima at $f_0 = \pm \sqrt{\beta /6 \alpha}$,
and the radius shrinks in the cycle again, responsible for the case (B).
The shrinking radius solution of the case (vi) ($f_0 = \pm \sqrt{3 \beta /2 \alpha}$) is rather particular.
That is related to the climbing hill motion,
starting from the point of the slope at the same level as the local maxima,
to the local maximum of the same side,
that needs infinite time to reach the summit.
A simple form of the time-dependent solution which is valid for arbitrary values of $f_0$ is shown in Appendix\,\ref{sec:A}.
For both (A) and (B) cases, the energy is evaluated as 
\begin{eqnarray}
E &=& \int \! d^3 x \mathrm{Tr} 
\left[
|\partial_t Z^A|^2 + \frac{4 \pi^2}{k^2} 
\left|
- \frac{k m}{2 \pi} Z^B + Z^A Z^{\dagger}_A Z^B - Z^B Z^{\dagger}_A Z^A 
\right|^2
\right] \nonumber \\
&=& \int \! d^3 x N (N-1) \left(
\dot{f}^2 + m^2 f^2 - \frac{4 \pi m}{k} f^4 + \frac{4 \pi^2}{k^2} f^6
\right) \nonumber \\
&=& N (N-1) 
\left(
\frac{4\pi^2}{k^2} f^6_0 - \frac{4 \pi m}{k} f^4_0 + m^2 f_0^2
\right)
\int \! d^3 x.
\end{eqnarray}
One can easily show that this expression is positive definite for all the 
value of $f_0$ except the critical points $f_0 = 0, \ \pm 
\sqrt{\frac{km}{2\pi}}$ at where the solution is given by $f = f_0 = \mathrm{const}.$

\section{Non-BPS solutions}
In this section, we analyze purely spatial, static solutions of the ABJM model.
These are generally non-BPS solutions but have interesting physical meaning.

A similar analysis in D-brane systems gives a hint to our case.
It is known that a BPS solution in the D-brane world-volume describes 
D-branes of lower dimensions.
Especially, $N$ monopole solution in D3-brane world-volume can be understood as 
$N$ D-strings attached to the D3-brane. 
In such D3$\perp$D1 configuration, the dual description by D-string 
world-volume effective theory is also possible \cite{CoMyTa}.
Multiple D-string world-volume effective theory is described by the 
non-abelian DBI action \cite{My} and generically a solution 
exhibits fuzzy configurations due to its noncommutative matrix structure.
The BPS solution of this D-strings effective theory 
is a fuzzy funnel whose cross-section is fuzzy $S^2$ and one 
divergent point along the D-string world-volume direction 
is interpreted as the location of a D3-brane that expands into the 
transverse direction of the D-strings.
In addition to this BPS solution 
it is known that there are non-BPS double funnel solutions 
whose cross-section is fuzzy $S^2$. They have
two divergent points within the finite segment along the D-string world-volume.
These divergent points are interpreted as the location of two separated parallel 3-branes.
According to how to choose the integration constant, there are two types 
of double funnel solutions. 
One is the wormhole-like solution describing D3 and anti-D3 connected by 
$N$ D-stings \cite{CaMa}. The other is the cusp solution studied in 
\cite{Ha} representing $N$ D-strings suspended by two D3-branes.
For large-$N$, the fuzziness of the configurations is 
effectively smoothed out and the energy and the R-R charge corresponding 
to the configurations precisely agrees both from the D3 and D1 point of views.

Analogous configurations are expected to exist also in eleven 
dimensional M-theory. 
Indeed, in the following, we demonstrate that such solutions actually 
exist in the ABJM model.

\subsection{M5/anti-M5 solution}
\begin{figure}[t]
\begin{center}
\includegraphics[scale=.7]{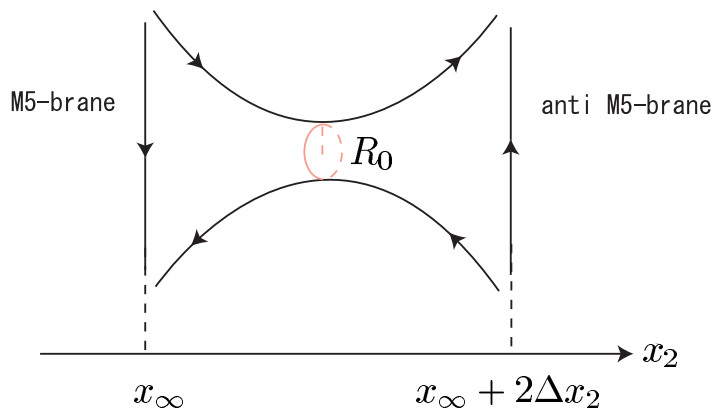}
\end{center}
\caption{A schematic picture of the solution (\ref{M5}), (\ref{anti-M5}).} \label{M5antiM5}
\end{figure}

Consider the ansatz $Z^A = f(x_2) S^A$ with a real function $f$. The 
massless equation of motion reduces to
\begin{eqnarray}
\partial_2^2 f (x_2) = 3 \alpha f^5 (x_2), \quad \alpha = \frac{4 \pi^2}{k^2}.
\label{double_funnel}
\end{eqnarray}
At fixed $x^2$, the configuration is interpreted as a fuzzy $S^3$.
The physical radius of the fuzzy $S^3$ is 
\begin{eqnarray}
R^2 = \frac{2 (N-1)}{T_2} f^2 (x_2).
\end{eqnarray}
Similar to the D3$\perp$D1 system 
in which the wormhole-like 
solutions connecting a D3 and an anti-D3 exist, there would be 
parallel M5-anti M5 solutions in the ABJM model.
The equation (\ref{double_funnel}) can be rewritten as
\begin{eqnarray}
 f' = \pm \sqrt{\alpha f^6 + c_1}, \label{plus_minus}
\end{eqnarray}
where $c_1$ is an integration constant.
Let us consider $c_1 \equiv - \alpha f^6_0, \ f_0 > 0$ case.
In this case, because $(f')^2 = \alpha f^6 - \alpha f^6_0 > 0$, 
the solution is valid only in the region $f > f_0$. 
Considering minus sign in (\ref{plus_minus}), $f$ is a 
monotonically decreasing function of $x_2$. Remarkably, $f' = 0$ at $f = 
f_0$ that happens within a finite value of $x_2$. 
The equation (\ref{plus_minus}) can be integrated to give 
\begin{eqnarray}
x_2  = x_{\infty} + \frac{1}{\sqrt{\alpha}} \int^{\infty}_{f} \frac{df}{\sqrt{ f^6 - 
f_0^6}}.
\end{eqnarray}
In terms of the physical radius, $x_2$ can be rewritten as
\begin{eqnarray}
x_2  = x_{\infty} + \frac{2 (N-1)}{\sqrt{\alpha} T_2} 
\int^{\infty}_R \! \frac{dR}{\sqrt{R^6 - R_0^6}}, \quad (R_0^2 = \frac{2 (N-1)}{T_2} f_0^2).
\label{M5}
\end{eqnarray}
At large $R$, this solution is approximated by the BPS fuzzy funnel 
solution, namely, an M5-brane at $x_2 = x_{\infty}$.
At $R = R_0 > 0, \ (x_2 = x_2 (R_0))$, the fuzzy funnel ceases to decreasing its radius. As in the case of D$p$/anti-D$p$ \cite{CaMa}, 
this solution can continue past 
this point by introducing another part of the solution
\begin{eqnarray}
x_2 = x_{\infty} + 2 \Delta x_2 - \frac{2 (N-1)}{\sqrt{\alpha} T_2} 
\int^{\infty}_R \frac{d R}{\sqrt{R^6 - R_0^6}}, \quad 
\Delta x_2 = x_2 (R_0) - x_{\infty}. \label{anti-M5}
\end{eqnarray}
The solution (\ref{anti-M5}) is smoothly connected to the solution 
(\ref{M5}) at $R=R_0$. The solution (\ref{anti-M5}) is the solution of 
(\ref{plus_minus}) for plus sign so that $R$ increases from $R_0$. 
For large $R$, the solution (\ref{anti-M5}) is approximated 
by another 5-brane at $x_2 = x_{\infty} + 2 \Delta x_2$.
As in the case of the D3$\perp$D1, 
this another 5-brane can be interpreted as an anti M5-brane.
This is because the orientation of the brane smoothly becomes opposite. 
See fig.\,\ref{M5antiM5} for a schematic picture.
The integral in the solution is evaluated as
\begin{eqnarray}
\int^{\infty}_R \frac{dR}{\sqrt{R^6 - R_0^6}} 
&=& \frac{1}{6 R^2_0} 
\left[
\int^{\infty}_1 \! d y \ y^{-\frac{5}{6}} (y-1)^{-\frac{1}{2}} 
- \int^{R^6/R^6_0}_1 \! d y \ y^{-\frac{5}{6}} (y-1)^{-\frac{1}{2}} 
\right] \nonumber \\
&=& \frac{1}{6 R^2_0} 
\left[
B \left( \frac{1}{2}, \frac{1}{3} \right)
- B \left( \frac{R^6 - R^6_0}{R^6}, \frac{1}{2}, \frac{1}{3} \right)
\right],
\end{eqnarray}
where we have changed the integration variable from $R$ to $y = R_0^{-6} 
R^6$.
Be careful that there is no point $x_2$ corresponding to $R = 0$.
From this expression, we have
\begin{eqnarray}
\Delta x_2 = x_2 (R_0) - x_{\infty}
= 2 (N-1) \frac{k}{2 \pi T_2} \frac{1}{6 R_0^2} B \left(\frac{1}{2}, 
\frac{1}{3} \right) \propto R_0^{-2}.
\end{eqnarray}
Note that this solution is not BPS but for $R_0 = 0$ limit, one of the two 5-branes goes to infinity 
and we recover the BPS fuzzy funnel solution which is nothing but an (anti) 
M5-brane.
For large separation of M5/anti-M5, $\Delta x_2 \to \infty$, 
the throat shrinks to zero-size. This is similar to fundamental strings 
between a D-brane and an anti-D-brane.

This fact can be confirmed from the M5-brane point of view. 
A scalar field $X$ in a single $p$-brane world-volume 
is governed by the Nambu-Goto action
\begin{eqnarray}
S_{p} = T_p \int \! d^{p+1} \xi \ \sqrt{1 + \partial_{\mu} X \partial^{\mu} X}.
\end{eqnarray}
Here $X$ describes fluctuation along a transverse direction to the 
$p$-brane and $T_p$ is a tension of the $p$-brane.
The static, spherically symmetric equation of motion of $X$ is  
\begin{eqnarray}
\frac{\partial}{\partial r} \left(\frac{r^{p-1} X'}{\sqrt{1 + 
(X')^2}}\right) = 0, \label{p-brane_eom}
\end{eqnarray}
where $r = \sqrt{\xi_1^2 + \cdots \xi_{p}^2}$ is a radial coordinate in the 
$p$-brane world-volume and the prime in the equation stands for 
the differentiation with respect to $r$.
The equation (\ref{p-brane_eom}) can be integrated to give
\begin{eqnarray}
X (r) = \int^{\infty}_r d r \frac{r_0^{p-1}}{\sqrt{r^{2p-2} - r_0^{2p-2}}}.
\end{eqnarray}
Here $r_0$ is an integration constant.
This precisely match with the solution (\ref{M5}) with $p=4$. 
Note that because the M5-brane shares the one-dimension with the M2-branes, 
$p = 5$ is effectively reduced to $p=4$. 

The energy of the solution is evaluated as 
\begin{eqnarray}
E = 
\frac{T_2^2}{4\pi} \frac{N}{N-1} \int \! dt dx^1 \left(\frac{2 
\pi^2}{k} \right) d R \frac{2 R^6 - R_0^6}{\sqrt{R^6 - R_0^6}}.
\end{eqnarray}
For large $N$ and large $R$ region, the energy is approximately given by
\begin{eqnarray}
E = T_5 \mathrm{vol (M5)},
\end{eqnarray}
where $T_5 \equiv \frac{T_2^2}{2\pi}$ and $\mathrm{vol (M5)} 
\equiv \int d x^1 \left(\frac{2 
\pi^2}{k} \right) R^3 d R$ are the tension and the volume of an M5-brane. 
Note that the factor $k$ in the volume appears due to the orbifolding $\mathbf{Z}_k$.
This is another evidence that the 5-branes located at the divergent 
points can be interpreted as an (anti) M5-brane. 
To clarify whether the brane we are considering is M5 or anti-M5, we need to 
evaluate its charge associated with 3-form in the eleven-dimensional supergravity.
Indeed, for the D-string case, we can show that the configuration of the 
double funnel becomes the source of the R-R 4-form potential and the 
opposite world-volume directions correspond to different signs of the 
charge \cite{CoMyTa}.
Unfortunately, in our case, we do not know the general coupling of 3-form in the 
multiple M2-brane world-volume theory and in order to clarify the issue more explicitly 
we need further study of this coupling to evaluate the charge 
corresponding to this configuration.
Currently, we are unable to determine this charge and this issue is beyond the scope of this 
paper.

\subsection{Cusp solution}
\begin{figure}[t]
\begin{center}
\includegraphics[scale=.7]{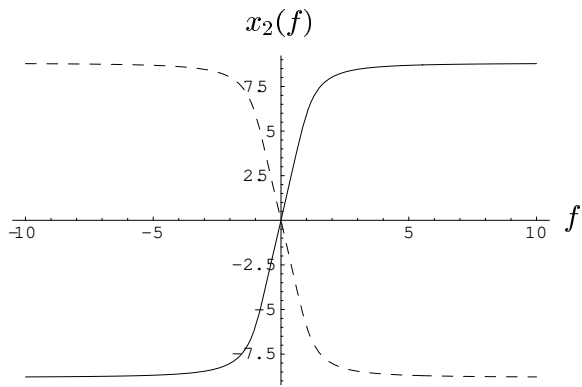}
\end{center}
\caption{An analytic plot of $ x_2 (f)$ with $k = f_0 = 1$. The solid and dashed 
lines correspond to plus and minus signs respectively in the equation (\ref{plus_minus}).} 
\label{cusp}
\end{figure}

Let us consider $c_1 \equiv  \alpha f^6_0, \ f_0 > 0$ case in the 
equation (\ref{plus_minus}).
In this case, the fuzzy funnel collapses at finite $x_2$ (the minus sign 
solution in (\ref{plus_minus})) where the fuzzy funnel is 
pinched off.
An analytic profile of the solution is plotted in fig.\,\ref{cusp}.

\begin{figure}[t]
\begin{center}
\includegraphics[scale=.8]{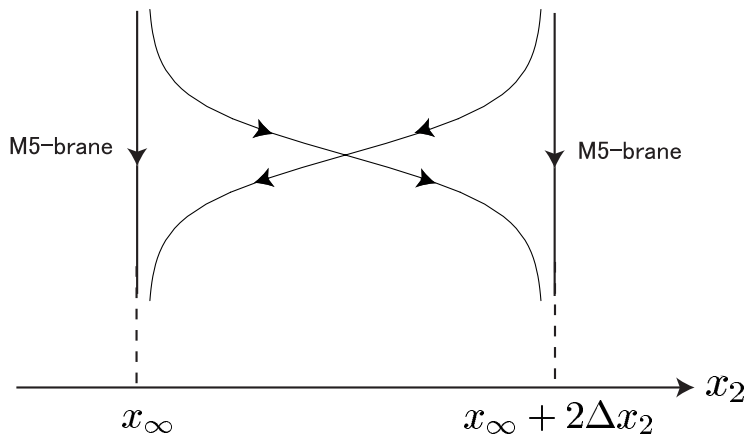}
\end{center}
\caption{A schematic picture of the cusp solution.} 
\label{cusp_sol}
\end{figure}

As discussed in \cite{CoMyTa, Ha} for D3$\perp$D1 case, there is no singularity at $R = 0$ 
and the solution can pass through this point smoothly provided the 
negative radius has physical meaning. 
Alternatively, once we want to keep the radius positive, the 
solution can be extended by introducing the solution for plus sign in (\ref{plus_minus}).
This is a cusp solution which would represent M2-branes connecting an M5-brane
and another M5-brane. This is because the orientation of the brane 
world-volume does not change.
A typical profile of this solution is illustrated in fig.\,\ref{cusp_sol}.
This situation is analogous to the one discussed in \cite{Ha} in the D3-brane context.
Especially, for $R> 0$ region, the integral can be evaluated as
\begin{eqnarray}
\int^{\infty}_{R} \frac{d R}{\sqrt{R^6 + R_0^6}}
&=& \frac{1}{6 R^2_0} 
\left[
\int^{\infty}_0 \! d y 
\ y^{- \frac{5}{6}} (y+1)^{-1/2} 
- \int^{R^6/{R^6_0}}_0 \! d y 
\ y^{- \frac{5}{6}} (y+1)^{-1/2} 
\right] \nonumber \\
&=& \frac{1}{6 R^2_0} 
\left[
B \left( \frac{1}{6}, \frac{1}{3} \right)
- B \left( \frac{R^6}{R^6 + R^6_0}, \frac{1}{6}, \frac{1}{3} \right)
\right].
\end{eqnarray}
Contrast to the M5/anti-M5 solution, there is a point $x_2$ corresponding to 
$R = 0$, -- a shrinking point of fuzzy $S^3$. The distance between two 
parallel M5-branes is given by
\begin{eqnarray}
\Delta x_2 = 2 (N-1) \frac{k}{2\pi T_2} \frac{1}{6 R_0^2} 
\left|
B \left(\frac{1}{6}, \frac{1}{3} \right) 
- B \left(\frac{1}{2}, \frac{1}{6}, \frac{1}{3} \right) 
\right| \propto R_0^{-2}.
\end{eqnarray}
A dual picture from the viewpoint of the M5-brane would be possible.
As in the case of the D3-brane analysis, 
turning on the self-dual gauge field in the M5-brane world-volume 
\cite{BaLeNuPaSoTo, AgPaPoSc} would play an important role to construct the cusp solution presented here.
Although there are two parallel M5-branes, this solution is not BPS. 
The BPS configuration is recovered in the limit $R_0 \to 0$ where 
one of the M5-brane goes to infinity. 

\subsection{Non-BPS solutions in mass deformed ABJM model}
Let us briefly discuss the static solution of the equation of motion in the mass deformed ABJM model. With the ansatz $Z^A = f(x_2) S^A,~f \in \mathbf R$, the equation of motion (\ref{m_eom}) can be rewritten as
\begin{eqnarray}
f'' = \frac{d V}{d f}, \hspace{10mm} V(f) \equiv \frac{2 \pi^2}{k^2} f^2 (f^2-v^2)^2, \hspace{10mm} v^2 \equiv \frac{k m}{2 \pi} = \frac{\beta}{2 \alpha}.
\end{eqnarray}
There is an integration constant $c$ defined by
\begin{eqnarray}
\frac{m^2 v^2}{2} c = \frac{1}{2} f'^2 - V(f).
\end{eqnarray}
Note that for $c=0$, this equation reduces to the BPS equation
\begin{eqnarray}
f' = \pm \frac{2 \pi}{k} f (f^2 - v^2),
\end{eqnarray}
and the solutions are given by $f_{\pm}$ in (\ref{massive_BPS}).
As in the case of the time-dependent solution, it is convenient to rewrite the equation of motion with respect to $\tau \equiv f^2$
\begin{eqnarray}
\tau'^2 
~=~ 8 f^2 [V(f) + c] 
~=~ \frac{4m^2}{v^4} \tau \left[ \tau (\tau-v^2)^2 + c v^6 \right],
\label{eq:tau_eom}
\end{eqnarray}
This equation can be simplified by using the following ansatz
\begin{eqnarray}
\tau(x) = \frac{3cv^2}{3h(m(x-x_0))-1}.
\end{eqnarray}
With this ansatz, the equation (\ref{eq:tau_eom}) reduces to the equation for the function $h$ 
\begin{eqnarray}
(h')^2 = 4 h^3 - g_2 h - g_3,
\label{eq:wp1}
\end{eqnarray}
where the parameters $g_2$ and $g_3$ are given by
\begin{eqnarray}
g_2 \equiv \frac{4}{3} \left( 6 c + 1 \right), \hspace{5mm} 
g_3 \equiv - 4 \left( c^2 + \frac{2}{3} c + \frac{2}{27} \right).
\end{eqnarray}
From equation (\ref{eq:wp1}), we find that the function $h$ is given by the Weierstrass's elliptic function $\wp$. Therefore, the static solution of the equation of motion is 
\begin{eqnarray}
\tau = f^2 = \frac{3cv^2}{3\wp(m(x-x_0))-1}.
\end{eqnarray}
For the values of the parameter $c<0$ and $c>0$, the solution has similar profiles as fig.\,\ref{M5antiM5} and fig.\,\ref{cusp_sol} respectively. By taking the limit $c \rightarrow 0$ with adjusting the parameter $x_0$ appropriately, we can reproduce the BPS solutions $f_{\pm}$ in (\ref{massive_BPS}).

\section{Conclusion and discussions}
In this paper, we have investigated several classical solutions 
of the ABJM model. 
The equation of motion for the scalar fields can be solved by the help of the BPS matrices. 
Due to the noncommutative nature of the matrices, the solutions 
generically exhibit fuzzy structures. 

In the first part of the main text, we studied 
time evolutions of fuzzy spheres
particularly focusing on fuzzy $S^3$s.
We found analytic solutions both for the massless and massive cases.
For the massless case, there is an oscillating fuzzy sphere solution 
which collapses into zero size then expands again.
The decay time is analytically derived.
For the massive case, there are solutions 
having different behaviors according to the value of the mass and 
the initial condition.
Remarkably, if the mass has an appropriate value, there are 
solutions that do not decay into zero size but keep its radius 
being positive.
This is different from the massless case where such solutions do not exist.

In the second half of the main body, we investigated purely spatial, 
static, generically non-BPS solutions. 
Similar to the D3$\perp$D1 system, we found two kinds of 
solutions.
One is the wormhole-like solution with its cross-section fuzzy $S^3$. 
This solution is given by smoothly connecting two 
fuzzy funnels. The two fuzzy funnels expand into two M5-branes 
with opposite world-volume direction. Therefore one of the M5-branes can 
be interpreted as an anti M5-brane.
The solution can be also analyzed from the viewpoint of dual M5-brane 
world-volume picture.
The other is the cusp solution connecting two 5-branes. 
In this solution, the fuzzy $S^3$ is 
pinched off at a point and then growing into the other 5-brane. 
Because the 5-brane world-volume direction does not change, the two 
5-branes are interpreted as two M5-branes.

There are a few comments on the study performed in this paper. 
The solutions examined in this paper are solutions of the effective 
action at leading order in 
the derivative expansion of a non-linear action.
This corresponds to the leading order Yang-Mills part in the expansion of the DBI action.
Actually, based on the novel higgs mechanism first proposed in 
\cite{MuPa}, when a scalar field develops a VEV and taking the large VEV 
and large $k$ limit with fixed $v/k$, the ABJM model reduces to the three-dimensional $\mathcal{N} = 
8$ $U(N)$ super Yang-Mills action -- the leading order of the effective action of $N$ D2-branes in 
type IIA string theory \cite{AhBeJaMa, PaWa}. 
In the discussion of D-brane world-volume effective action, 
such a leading order solution (including the wormhole and cusp solutions) 
frequently gives a solution in the full non-linear level and the 
solutions found in this paper would keep being solutions at the full 
non-linear M2-brane effective action.
Some non-linear version of multiple M2-brane effective theories were proposed 
in \cite{IeRu2, AlMu, Ga}. 

Another important point is to find M2-brane effective action in general supergravity background. 
The ABJM model is interpreted as an multiple M2-brane effective action 
in flat space-time $\mathbf{C}^4/\mathbf{Z}_k$, and couplings with the background supergravity 
field cannot be described correctly in this model. It is important to find the correct 
coupling of these supergravity fields in the multiple M2-branes.

In this paper, we analyzed the simplest class of the solutions, $Z^A 
\not= 0$, $W_A = A_{\mu} = \hat{A}_{\mu} = 0$. 
It would be interesting to investigate the effects of non-zero gauge fields 
like the situation studied in 
\cite{CoMyTa} where $(p,q)$-strings attached to a D3-brane.

\subsection*{Acknowledgments}
The work of T.~F. is supported
by the Research Fellowships of the Japan Society for the Promotion of 
Science (JSPS) for Young Scientists. 
The work of K.~I is supported by Iwanami Fujukai Foundation.
S.~S. would like to thank K.~Hanaki, S.~Kawai and S.~Terashima for 
useful discussions and conversations, and especially C.~Montonen for 
reading the manuscript.
The work of S.~S. is supported by bilateral exchange program between 
Japan Society for the Promotion of Science (JSPS) and the Academy of 
Finland (AF).

\begin{appendix}
\section{BPS matrix}
The BPS matrices $S^A \ (A = 1,2)$ satisfy the following relations.
\begin{eqnarray}
& & S^A = S^B S^{\dagger}_B S^A - S^A S^{\dagger}_B S^B, \\
& & S^{\dagger}_A = S^{\dagger}_A S^B S^{\dagger}_B 
- S^{\dagger}_B S^B S^{\dagger}_A.
\end{eqnarray}
The explicit form of the matrices $S^A$ was first found in 
\cite{GoRoRaVe}. This is given by
\begin{eqnarray}
& & (S^{\dagger}_1)_{mn} = \sqrt{m -1} \delta_{mn}, \quad 
(S^{\dagger}_2)_{mn} = \sqrt{N-m} \delta_{m+1,n}, \\
& & S^1 S^{\dagger}_1 = \mathrm{diag} (0,1,2, \cdots, N-1) = 
S^{\dagger}_1 S^1, \\
& & S^2 S^{\dagger}_2 = \mathrm{diag} (0,N-1,N-2, \cdots, 1), \\
& & S^{\dagger}_2 S^2 = \mathrm{diag} (N-1, N-2, \cdots, 1,0), \\
& & S^A S^{\dagger}_A = \mathrm{diag} (0, N, \cdots, N).
\end{eqnarray}
From this expression, we have
\begin{eqnarray}
\mathrm{Tr} S^A S^{\dagger}_A = \mathrm{Tr} S^{\dagger}_A S^A = N (N-1).
\end{eqnarray}
Let us see the linearization of the matrices $S^A$ 
in the right hand side of the massless equation of motion (\ref{eom}).
Assuming an ansatz $Z^A = f (x) S^A$, 
the matrix structure on the right hand side of (\ref{eom}) is
\begin{eqnarray}
3 (S^B S^{\dagger}_B)^2 S^A + 3 S^A (S^{\dagger}_B S^B)^2 
- 2 S^A S^{\dagger}_B (S^C S^{\dagger}_C ) S^B - 2 (S^C S^{\dagger}_C) 
S^A (S^{\dagger}_B S^B) - 2 S^B (S^{\dagger}_C S^C) S^{\dagger}_B S^A
\equiv M^A.
\nonumber \\
\end{eqnarray}
The BPS matrices satisfy the relation
\begin{eqnarray}
(S^B S^{\dagger}_B)^2 S^A &=& (S^B S^{\dagger}_B) S^A + (S^B S^{\dagger}_B) S^A
(S^{\dagger}_C S^C), \\
S^A (S^{\dagger}_B S^B)^2 &=& 
- S^A (S^{\dagger}_B S^B) + (S^B S^{\dagger}_B) S^A (S^{\dagger}_C S^C).
\end{eqnarray}
Therefore
\begin{eqnarray}
3 (S^B S^{\dagger}_B)^2 S^A + 3 S^A (S^{\dagger}_B S^B)^2 
= 3 S^A
+ 6 (S^B S^{\dagger}_B) S^A (S^{\dagger}_C S^C).
\end{eqnarray}
Then
\begin{eqnarray}
M^A = 3 S^A + 4 (S^B S^{\dagger}_B) S^A (S^{\dagger}_C S^C) 
- 2 S^A S^{\dagger}_B (S^C S^{\dagger S_C}) S^B 
- 2 S^B (S^{\dagger}_C S^C) S^{\dagger}_B S^A.
\end{eqnarray}
Using the relation
\begin{eqnarray}
S^{\dagger}_B S^B &=& S^{\dagger}_B (S^C S^{\dagger}_C) S^B - 
 (S^{\dagger}_C S^C)^2, \\
 S^B S^{\dagger}_B &=& (S^C S^{\dagger}_C)^2 - S^B (S^{\dagger}_C S^C) S^{\dagger}_B,
\end{eqnarray}
we have
\begin{eqnarray}
& & 4 (S^B S^{\dagger}_B) S^A (S^{\dagger}_C S^C) 
- 2 S^A S^{\dagger}_B (S^C S^{\dagger S_C}) S^B 
- 2 S^B (S^{\dagger}_C S~C) S^{\dagger}_B S^A \nonumber \\
&=& 4 (S^B S^{\dagger}_B) S^A (S^{\dagger}_C S^C) 
- 2 S^A (S^{\dagger}_B S^B) + 2 (S^B S^{\dagger}_B) S^A + 2 S^A 
(S^{\dagger}_B S^B) \nonumber \\
& & \qquad - 2 (S^B S^{\dagger}_B) S^A (S^{\dagger}_C S^C)
- 2 (S^B S^{\dagger}_B) S^A - 2 (S^B S^{\dagger}_B) S^A (S^{\dagger}_C 
S^C) \nonumber \\
&=& 0.
\end{eqnarray}
Finally, we find
\begin{eqnarray}
M^A = 3 S^A.
\end{eqnarray}
For the massive case, there is only an additional term coming from the 
second term in the right hand side in (\ref{massive_eom}). 
This can be trivially linearized by using the relation (\ref{BPS_matrix_condition}).

\section{Moving fuzzy funnels and a domain wall}
As discussed in section 1, the equation of motion is 
world-volume Lorentz invariant. Therefore one way to obtain 
a non-trivial time-dependent solution is Lorentz boosting of 
known purely spatial solutions. 
In this appendix, we show that 
moving fuzzy funnels and domain wall solutions with constant velocity 
are obtained by boosting known BPS solutions.
Consider an ansatz
\begin{eqnarray}
Z^A (x) = f(x) S^A, \quad f \in \mathbf{R}, \label{ansatz}
\end{eqnarray}
then the equation of motion for massless case (\ref{eom}) reduces to
\begin{eqnarray}
(-\partial_t^2 + \partial_1^2 + \partial_2^2) f (x) S^A = \frac{12\pi^2}{k^2} f^5 (x) S^A. 
\end{eqnarray}
The BPS solution 
\begin{eqnarray}
f (x) = f (x_2) = \sqrt{\frac{k}{4 \pi}} (x_2 + x_0)^{-\frac{1}{2}}
\label{BPS_massless}
\end{eqnarray}
satisfies the equation (\ref{eom}). In addition to the static solution, 
we can explicitly show the following solution satisfies the equation,
\begin{eqnarray}
f(x_2, t) = \sqrt{\frac{k}{4\pi}} \left( \gamma x_2 \pm \sqrt{\gamma^2 - 1} 
				 t \right)^{- \frac{1}{2}}.
\label{boosted_M5}
\end{eqnarray}
Here $0 < \gamma \le 1$ is a real parameter.
This is just the Lorentz boost of the BPS solution (\ref{BPS_massless}) 
with 
\begin{eqnarray}
x^{\prime 2} = \gamma (x^2 - v t), \quad \gamma = \frac{1}{\sqrt{1 - v^2/c^2}}.
\label{Lorentz_boost}
\end{eqnarray}
This solution would represent moving an M5-brane with constant velocity $v = 
\pm c \frac{\sqrt{\gamma^2 - 1}}{\gamma}$. $c$ is the speed of light.
Let us check this fact. The cross-section of the solution 
(\ref{boosted_M5}) at fixed $x_2, t$ is a fuzzy $S^3$ with radius
\begin{eqnarray}
R^2 = \frac{k (N-1)}{2 \pi T_2} \frac{1}{\alpha x_2 \pm \sqrt{\alpha^2 - 1} t}.
\end{eqnarray}
Then the energy of the configuration (\ref{boosted_M5}) 
is evaluated as
\begin{eqnarray}
E &=& \int d t d x^1 d x^2 \mathrm{Tr} 
\left[
|\partial_t Z^A|^2 + |\partial_2 Z^A|^2 + 
\frac{4 \pi^2}{k^2} |Z^A Z^{\dagger}_A Z^B - Z^B Z^{\dagger}_A Z^A|^2
\right] \nonumber \\
&=& \gamma \frac{T_2^2}{2 \pi} \frac{N}{N-1} \int dt d x^1 \left( 
\frac{2 \pi^2}{k} \right) R^3 d R.
\label{energy}
\end{eqnarray}
At large-$N$, this reduces to
\begin{eqnarray}
E = \frac{Mc^2}{\sqrt{1-v^2/c^2}}
\quad (N \to \infty),
\end{eqnarray}
where 
\begin{eqnarray}
M = T_5 \mathrm{vol} (M5)
\end{eqnarray}
is the mass of an M5-brane.
This energy precisely match with the one of an M5-brane 
moving with velocity $v < c$. 
Similar to the massless case, we obtain moving deformed fuzzy funnel and 
domain wall by Lorentz boosting known massive BPS solutions
\begin{eqnarray}
f_{\pm} (x_2, t) = 
\sqrt{\frac{k m}{2 \pi}} 
\left(
1 \pm e^{-2 m x'_2}
\right)^{- \frac{1}{2}},
\end{eqnarray}
where $x'$ is given by the equation (\ref{Lorentz_boost}).

\section{Time-dependent solution in a simple form}\label{sec:A}
In this section, we derive a simple form of the time-dependent solution which is valid for arbitrary values of $f_0$. 
The equation of motion (\ref{time_massive}) can be rewritten as 
\begin{eqnarray}
\ddot f = - \frac{d V}{d f}, \hspace{10mm} V(f) \equiv \frac{2 \pi^2}{k^2} f^2 (f^2-v^2)^2, \hspace{10mm} v^2 \equiv \frac{k m}{2 \pi} = \frac{\beta}{2 \alpha}.
\end{eqnarray}
This is nothing but the equation of motion for the particle in the potential $V(f)$. Therefore we can define the conserved energy as
\begin{eqnarray}
E \equiv \frac{1}{2} \dot f^2 + V(f).
\end{eqnarray}
If we assume that $f(t_0) = f_0$ and $\dot f(t_0) = 0$, the energy is given by
\begin{eqnarray}
E = V(f_0).
\end{eqnarray}
The equation (\ref{massive_osc_eq}) ($\dot \tau^2 = - 8 f^2 [V(f)-V(f_0)]$) can be cast into a simple form by using the following ansatz
\begin{eqnarray}
\tau(t) = f_0^2 \left(1 - \frac{a}{b+h(m(t-t_0))} \right), \hspace{5mm} a \equiv \frac{(f_0^2-v^2)(3f_0^2-v^2)}{v^4}, \hspace{5mm} b \equiv \frac{1}{3} + \frac{2f_0^2(f_0^2-v^2)}{v^4}.
\end{eqnarray}
Then, the equation for the function $h$ is given by
\begin{eqnarray}
\dot h^2 = 4 h^3 - g_2 h - g_3,
\label{eq:wp}
\end{eqnarray}
where the parameters $g_2$ and $g_3$ are given by
\begin{eqnarray}
g_2 \equiv - \frac{4}{3} \left( 2 c - 1 \right), \hspace{5mm} 
g_3 \equiv \frac{4}{27} \left( 3 c^2 - 6 c + 2 \right), \hspace{5mm}
c   \equiv \frac{3 f_0^2(f_0^2-v^2)^2}{v^6}.
\end{eqnarray}
The solution to the equation (\ref{eq:wp}) is nothing but the equation for the Weierstrass's elliptic function $\wp$. Therefore, the time-dependent solution for $\tau$ is given by
\begin{eqnarray}
\tau = f^2 =f_0^2 \left(1 - \frac{a}{b+\wp(m(t-t_0))} \right).
\end{eqnarray}
This solution is valid for arbitrary value of $f_0$.

\end{appendix}

\end{document}